\documentclass[amsmath,amssymb,aps,prd,a4paper,onecolumn,notitlepage]{revtex4-1}
\usepackage{graphicx}
\usepackage{subfigure}
\usepackage{amsmath}
\usepackage[section]{placeins}
\usepackage{slashed}
\usepackage{color}
\usepackage{soul}
\usepackage{appendix}
\usepackage{placeins}
\usepackage{float}
\usepackage[utf8]{inputenc}
\usepackage{pstricks}
\usepackage{multirow}
\usepackage{longtable}
\usepackage{hyperref}
\usepackage{subfigure}
\usepackage{epsfig}
\usepackage{braket}
\usepackage[normalem]{ulem} 
\usepackage{longtable,booktabs}
\newcommand{\be}{\begin{equation}}
\newcommand{\ee}{\end{equation}}
\newcommand{\ben}{\begin{eqnarray}}
\newcommand{\een}{\end{eqnarray}}

\usepackage{stackengine}

\usepackage{verbatim}
\usepackage{hyperref}
\usepackage{bm}
\usepackage{tikz}
\usetikzlibrary{tikzmark}

\begin{document}

%\title{Study of the lineshape of $D_{1}(2420)$ and  $D_{1}(2430)$ from $B^{-}\to D^{*+}\pi^{-}\pi^{-}$}

%=========================================================================&
%=========================================================================&
\title{$J/\psi \Lambda $ femtoscopy and the nature of  $
P_{\psi s}^{\Lambda}(4338)$}

\author{Yi-Bo Shen}
\affiliation{School of Physics, Beihang University, Beijing 102206, China}

\author{Zhi-Wei Liu}
\affiliation{School of Physics, Beihang University, Beijing 102206, China}

\author{Ming-Zhu Liu}
\email[Corresponding author: ]{liumz@lzu.edu.cn}
\affiliation{
Frontiers Science Center for Rare isotopes, Lanzhou University,
Lanzhou 730000, China}
\affiliation{ School of Nuclear Science and Technology, Lanzhou University, Lanzhou 730000, China}

\author{Rui-Xiang Shi}
\affiliation{Department of Physics, Guangxi Normal University, Guilin 541004, China}
\affiliation{Guangxi Key Laboratory of Nuclear Physics and Technology, Guangxi Normal University, Guilin 541004, China}

\author{Chu-Wen Xiao}
\affiliation{Department of Physics, Guangxi Normal University, Guilin 541004, China}
\affiliation{Guangxi Key Laboratory of Nuclear Physics and Technology, Guangxi Normal University, Guilin 541004, China}

\author{Wei-Hong Liang}
%\email[Corresponding author: ]{liangwh@mailbox.gxnu.edu.cn}
\affiliation{Department of Physics, Guangxi Normal University, Guilin 541004, China}
\affiliation{Guangxi Key Laboratory of Nuclear Physics and Technology, Guangxi Normal University, Guilin 541004, China}
 
\author{Li-Sheng Geng}
\email[Corresponding author: ]{lisheng.geng@buaa.edu.cn}
\affiliation{School of
Physics,  Beihang University, Beijing 102206, China}
\affiliation{Sino-French Carbon Neutrality Research Center, \'{E}cole Centrale de P\'{e}kin/School
of General Engineering, Beihang University, Beijing 100191, China}
\affiliation{Peng Huanwu Collaborative Center for Research and Education, Beihang University, Beijing 100191, China}
\affiliation{Beijing Key Laboratory of Advanced Nuclear Materials and Physics, Beihang University, Beijing 102206, China }
\affiliation{Southern Center for Nuclear-Science Theory (SCNT), Institute of Modern Physics, Chinese Academy of Sciences, Huizhou 516000, China}

%\affiliation{Instituto de Física, Universidade Federal da Bahia, Campus Ondina, Salvador, Bahia 40170-115, Brasil}

%=========================================================================&
%=========================================================================&
%=========================================================================&
\begin{abstract}

Over the past two decades, numerous exotic hadron states have been discovered, yet their underlying nature remains unclear. It is widely acknowledged that understanding hadron-hadron interactions is essential to unraveling their properties. Hadron spectroscopy is a powerful tool for this endeavor, providing rich experimental data that can shed light on exotic systems. Recently, the LHCb experiment analyzed the process $B^{-} \rightarrow J/\psi \Lambda \bar{p}$ and observed a narrow peak in the $J/\psi \Lambda$ invariant mass spectrum, regarding it as a candidate for a pentaquark. In this work, we extract the coupled-channel $J / \psi \Lambda-\bar{D} \Xi_c-\bar{D}_s \Lambda_c$ potential based on the $J/\psi \Lambda$ invariant mass spectrum. Our results indicate the existence of a bound state below the $\bar{D} \Xi_c$ mass threshold, corresponding to the experimentally measured state $P_{cs}(4338)$.   Furthermore, we predict the scattering lengths and momentum correlation functions for the $J/\psi \Lambda$ and $\bar{D}\Xi_c$ channels, which serve as theoretical references for future femtoscopy experiments.

\date{\today}
\end{abstract}
%=========================================================================&
%=========================================================================&
%=========================================================================&

\maketitle

%=========================================================================&
%=========================================================================&
%=========================================================================&
\section{\label{sec:intro}Introduction}

Quantum Chromodynamics (QCD), the fundamental theory of the strong interaction, exhibits weak coupling at high energies—where perturbative methods based on quarks and gluons are applicable—but becomes strongly coupled in the low-energy regime. In this domain, the relevant degrees of freedom are hadrons rather than quarks and gluons, a phenomenon known as color confinement. This transition makes it challenging to study the strong interaction at the quark level in low-energy processes. As a result, low-energy strong interactions remain difficult to determine both theoretically and experimentally. Hadron spectroscopy plays a vital role in probing strong interactions across energy scales. A typical example is the study of charmonium spectra, which has verified the behavior of strong interactions in both low- and high-energy regimes~\cite{Eichten:1974af}. In recent years, an increasing number of hadrons beyond the conventional hadrons (mesons made by a quark and an antiquark, baryons made by three quarks)~\cite{Gell-Mann:1964ewy} have been observed~\cite{Brambilla:2010cs,Olsen:2017bmm,Brambilla:2019esw}, indicating that non-perturbative QCD effects, such as hadron-hadron interactions and three-body hadron forces~\cite{Wu:2025fzx,Pan:2025vyg}, play an important role in explaining the properties of these states~\cite{Chen:2016qju,Lebed:2016hpi,Oset:2016lyh,Esposito:2016noz,Dong:2017gaw,Guo:2017jvc,Ali:2017jda,Karliner:2017qhf,Guo:2019twa,Liu:2024uxn,Husken:2024rdk,Wang:2025sic,Doring:2025sgb}. As a result, the study of hadron-hadron interactions has become one of the central topics in hadron physics.

Invariant mass spectrum encodes essential information about hadron-hadron interactions~\cite{Dong:2020hxe}, making them one of the primary tools for extracting such interactions.   
Recently, femtoscopy—a technique that analyzes momentum correlations between particles produced in high-energy collisions—has emerged as a powerful and complementary tool for probing hadron-hadron interactions~\cite{Fabbietti:2020bfg,Liu:2024uxn,Molina:2023oeu,Liu:2025rci,Ge:2025put}. By measuring momentum correlation functions (CFs), femtoscopy offers valuable insights into hadron-hadron interactions, enabling the discrimination between strongly attractive, weakly attractive, and repulsive hadron-hadron forces~\cite{Liu:2023uly,Liu:2023wfo}.  In addition, femtoscopy can help distinguish whether a particular exotic state is bound, virtual, or resonant, which corresponds to various strengths of the underlying hadron-hadron interaction~\cite{Liu:2024nac,Liu:2025wwx,Liu:2025nze}.  In particular, cusp effects arising from coupled-channel interactions have been investigated through momentum CFs~\cite{Kamiya:2019uiw,Liu:2022nec,Xie:2025xew}.   Furthermore, we found that momentum CFs have the potential to study the couplings of bare QCD states to hadron-hadron interactions~\cite{Shen:2025qpj}. As indicated in Ref.~\cite{Kievsky:2023maf,ALICE:2025aur}, momentum CFs can also be used to study the three-body interaction among three nucleons.

 In this work, we focus on the interactions between the $J/\psi$ meson and light baryons, while the underlying mechanisms remain controversial. Unlike open-charm hadron interactions, those between  $J/\psi$  and light baryons are suppressed by the Okubo-Zweig-Iizuka(OZI) rule and are difficult to understand by the meson exchange mechanisms. The interaction between the $J/\psi$ and a nucleon has been studied in a lot of theoretical work.  The method of multipole expansion and low-energy QCD theorems were used to calculate the $c\bar{c}$-nucleon scattering amplitude, predicting a scattering length $a_{J/\psi p}=-0.05$ fm~\cite{Kaidalov:1992hd}, which was later improved in the same approach, i.e., $a_{J/\psi p}<-0.37$~fm~\cite{Sibirtsev:2005ex}.  The  QCD sum rule analysis calculated the  $J/\psi N$  scattering length $a_{J/\psi p}=-0.1$ fm~\cite{Hayashigaki:1998ey}. Later, based on the experimental data for hidden charm photo-production on a proton as input, the vector meson dominance model estimated the scattering length $a_{J/\psi p}=-0.046$ fm~\cite{Gryniuk:2016mpk}, $a_{J/\psi p}=-0.003\sim -0.025$ fm~\cite{Pentchev:2020kao}, and $a_{J/\psi p}=-0.004$ fm~\cite{Wang:2022xpw}.   Du et al. estimated the  $J/\psi p$ scattering lengths through the coupled-channel mechanism~\cite{Du:2020bqj}, and later estimated them through the soft-gluon mechanism $a_{J/\psi p}<-0.16$ fm~\cite{Wu:2024xwy}.   
Very recently, lattice QCD simulations of the $J/\psi p$ interaction found a scattering length around $a_{J/\psi p}=-0.3$ fm~\cite{Lyu:2024ttm}.

Nevertheless, it is necessary to study the $J/\psi p$ interaction as well as the $J/\psi \Lambda$ interaction using alternative physical observables. In this work, 
  we employ the contact range effective field theory(EFT) coupled-channel potential, including both open-charm and hidden-charm channels, to study both $J/\psi p$ and $J/\psi \Lambda$ interactions, where the unknown parameters are determined by the $J/\psi p$~\cite{LHCb:2015yax,LHCb:2019kea} and $J/\psi \Lambda$~\cite{LHCb:2022ogu}  invariant mass spectra in b-flavored hadron decays by the LHCb Collaboration.  Using the obtained potentials, we explore resonant structures near the thresholds of open-charm channels and confirm the nature of the pentaquark states discovered by the LHCb Collaboration. Then we calculate their corresponding CFs and scattering lengths. In Ref.~\cite{Liu:2025oar}, we employed the lattice QCD $J/\psi p$ potentials to predict the $J/\psi p$ momentum CFs. Therefore, we only focus on the $\bar{D}\Xi_c$ and $J/\psi\Lambda$ CFs in this work.  Our predictions may help to determine interactions between charmonia and light baryons.

In this paper, we first present the formulae for calculating the CFs and scattering lengths of a pair of hadrons in Section II. Then,  using the potentials determined by fitting the  $J/\psi\Lambda$ invariant mass spectrum in the decay $B^{-} \rightarrow J/\psi \Lambda \bar{p}$, we predict the CFs and scattering lengths of $J/\psi\Lambda$ and $\bar{D}\Xi_c$ systems.  In addition, based on the potentials by fitting the  $J/\psi p$ invariant mass spectrum in the decay $ \Lambda_b \rightarrow J/\psi  p K$, we also predict the scattering lengths of $J/\psi p$ and $\bar{D}^{(*)}\Lambda_c$  systems.    Finally, a summary is given in the last section.

\section{Theoretical Formalism}

In this section, we outline the theoretical framework for calculating CFs. Based on the Koonin-Pratt(KP) formula~\cite{Koonin:1977fh,Pratt:1990zq}, CFs are determined by two key ingredients:    1) Particle-Emitting Source: Characterizing the spatial distribution of hadron emissions in relativistic heavy-ion collisions; 2) Scattering Wave Function: Encoding the final-state interactions between hadron pairs and derived from reaction amplitude $T$-matrices. The latter provides direct access to the hadron-hadron interaction of the system under study.  

First, we present the CFs based on the KP formula in the coupled channel case~\cite{Liu:2024nac}:

\begin{equation}
\begin{aligned}
C_j\left(k_j\right) & =1+\int_0^{\infty} \mathrm{d}^3  \vec{r} S_{12}\left[\sum_i \nu_i\left|\delta_{i j} j_0\left(k_j r\right)+T_{i j} \widetilde{G}_i\right|^2-\left|j_0\left(k_j r\right)\right|^2\right],
\label{cf}
\end{aligned}
\end{equation}
where $\nu_i$ is the weight for each component of the multi-channel wave function, and the sum runs over all possible coupled channels. For simplicity, we assume the weights are equal to $1$ in this study. $S_{12}$   is characterized by a  Gaussian source function  $S_{12}(r)=\text{Exp}~[-r^{2}/(4R^2)]/(2\sqrt{\pi}R)^3$ with a single parameter $R$,  $j_{0}(kr)$ is the spherical Bessel function, and  $k=\sqrt{s-(m_1+m_2)^2}\sqrt{s-(m_1-m_2)^2}/(2\sqrt{s})$ represents the center-of-mass (c.m.) momentum of the particle pair with the masses being  $m_1$ (meson) and $m_2$ (baryon) and c.m. energy $\sqrt{s}$.  The $\widetilde{G}$ is given by

\begin{equation}
\widetilde{G}(r, \sqrt{s})=\int_0^{q_{\text {max }}} \frac{\mathrm{d}^3 \vec{k}^{\prime}}{(2 \pi)^3} \frac{\omega_1(k')+\omega_2(k')}{2 \omega_1(k')\, \omega_2(k')} \frac{2m_2\cdot j_0\left({k^{\prime}} r\right)}{s-\left[\omega_1(k')+\omega_2(k')\right]^2+i \varepsilon}
\end{equation}

with $\omega_{i}(k')=\sqrt{m_i^2+k^{\prime2}}$. The scattering matrix $T$ is obtained by solving the Bethe-Salpeter equation
\begin{equation}
    T=[1-VG]^{-1}V,
\label{T}
\end{equation}
where $V$ is the coupled-channel potential described by the contact-range EFT and $G(\sqrt{s})$ is the loop function
\begin{equation}
\begin{split}
G(\sqrt{s})&=\int_{0}^{q_\mathrm{max}}\frac{{\rm d}q~q^{2}}{4\pi^{2}}\frac{\omega_1(q)+\omega_2(q)}{\omega_1(q)\cdot\omega_2(q)}
    \frac{2m_2}{s-[\omega_(q)+\omega_2(q)]^{2}+i\varepsilon}.
\end{split}
\label{loop function}
\end{equation}
In this work, we employ the momentum cutoff approach to regularize the divergent loop function, setting $q_{max} = 1$~GeV. The unknown couplings of the coupled-channel potentials are determined by fitting to the relevant experimental invariant mass spectrum. The amplitude of the final state $J / \psi \Lambda$ in the decay process of $J/\psi \Lambda$ in $B^{-} \rightarrow J / \psi \Lambda \bar{p}$ is expressed as 
\begin{equation}
    T_{B}=d+dG_{J/\psi \Lambda}T,
    \label{T0}
\end{equation}
where the term $d$ represents the contact term of the three-body weak decay $B^{-} \rightarrow J / \psi \Lambda \bar{p}$, and the term $dG_{J/\psi \Lambda}T$ represents the $J/\psi \Lambda$ final-state interaction.  Here, we focus exclusively on the $J/\psi \Lambda$ invariant mass spectrum, without considering the production mechanism of the three-body weak decay $B^{-} \rightarrow J / \psi \Lambda \bar{p}$.  Such a weak-interaction vertex arises from short-range interactions, whereas long-range interactions govern final-state interactions. According to the factorization ansatz, the long-range interactions remain unaffected by short-range contributions.     As a result, the production rate of the decay $B^{-} \rightarrow J / \psi \Lambda \bar{p}$ is parameterized as a constant $d$, which is determined by fitting experimental data.  With the total amplitude $T_B$,  one can obtain the $J/\psi \Lambda$ invariant mass spectrum of the decay $B^{-} \rightarrow J / \psi \Lambda \bar{p}$  ~~\cite{Burns:2022uha} 
\begin{equation}
    \frac{d\Gamma}{d M_{inv}}=\frac{\sqrt{\lambda(m_{B}^{2},s,m_{\bar{p}}^{2})} \sqrt{\lambda(s,m_{J/\psi }^{2},m_{\Lambda}^{2})}}{4m_{B} \sqrt{s}}\times \left|T_B\right|^{2},
    \label{M}
\end{equation}
where $\lambda(a,b,c)=a^2+b^2+c^2-2ab-2ac-2bc$ and the constant factor of $1/(2\pi)^{3}$ and $1/4m_B^{2}$ in phase space is absorbed into the weak-interaction parameter. 

Once the couplings in the potentials are determined, one can further derive  the  scattering amplitude $T_{ii}$ of channel $i$,  then the corresponding  scattering length $a$  at the mass threshold is given by~\cite{Ikeno:2023ojl,Shen:2024npc},
\begin{equation}
    -\frac{1}{a_i}=-\left.\frac{8 \pi \sqrt{s}}{2m_{i2}} T_{ii}^{-1}\right|_{s=s_{\mathrm{th}}},
\end{equation}
where $\sqrt{s_{th}}=(m_1+m_2)$ represents the mass threshold. %And the $g$ .

\section{Results and discussion}

In this section, we extract the corresponding potentials in two scenarios by fitting the $J/\psi \Lambda$ invariant mass spectrum: scenario I (two-channel system $J/\psi \Lambda - \bar{D} \Xi_c$) and scenario II (three-channel system $J/\psi \Lambda - \bar{D} \Xi_c - \bar{D}_s \Lambda_c$).  Subsequently, we calculate the corresponding scattering lengths and CFs. 

For scenario I, the $J/\psi \Lambda - \bar{D} \Xi_c$  potential     is parameterized as  a symmetric matrix
\begin{equation}
\mathbf{V} =\left(\begin{array}{ll}
a & c \\
c & b
\end{array}\right),
\label{v1}
\end{equation}
where $a$,  $b$, and  $c$ characterize the scattering processes  $J/\psi \Lambda \to J/\psi \Lambda$, $\bar{D} \Xi_c  \to \bar{D} \Xi_c$, and  $\bar{D} \Xi_c  \to J/\psi \Lambda$, respectively.   
The corresponding loop functions are given by a diagonal matrix
\begin{equation}
    G=\left(\begin{array}{cc}
G_{J/\psi \Lambda} & 0 \\
0 & G_{\bar{D}\Xi_c}
\end{array}\right). 
\end{equation}
Using the loop function $G$ and potential $V$, the scattering matrix $T$ can be obtained from Eq.~\eqref{T}. In this work, the scattering matrix $T$ describes the final-state interactions of the $B$ meson decays. For scenario I, in addition to the direct decay $B^{-} \rightarrow J / \psi \Lambda \bar{p}$, the decay $B^{-} \rightarrow  \bar{D}\Xi_c \bar{p}$ can transit into $B^{-} \rightarrow J / \psi \Lambda \bar{p}$ via the final-state interaction $\bar{D} \Xi_c  \to J/\psi \Lambda$. The weak decays $B^{-} \rightarrow J / \psi \Lambda \bar{p}$ and $B^{-} \rightarrow  \bar{D}\Xi_c \bar{p}$ are characterized by two unknown couplings $d$ and $f$, respectively. As a result, the decay of $B^{-} \rightarrow  J/\psi \Lambda \bar{p}$ with final-state interactions is written as $T_B=d+dG_{J/\psi \Lambda}T_{11}+fG_{\bar{D}\Xi_{c}}T_{21}$. Following Ref.~\cite{Burns:2022uha}, the background is parameterized by a constant $x_{back}$. Finally, the total amplitude $T_{B}$ is given by 
\begin{equation}
    T_{B}=x_{back}+d+dG_{J/\psi \Lambda}T_{11}+fG_{\bar{D}\Xi_{c}}T_{21}.
    \label{Tb}
\end{equation}

For scenario II, the coupled-channel $J / \psi \Lambda-\bar{D} \Xi_c-\bar{D}_s \Lambda_c$ potential $\mathbf{V}^{\prime}$ is parameterized as   
\begin{equation}
\mathbf{V}^{\prime}=\left(\begin{array}{ccc}
a & c & x \\
c &b & y \\
x & y &z
\end{array}\right),
\label{V2}
\end{equation}
where $x$, $y$, and $z$ characterize the potentials  of  $\bar{D}_s\Lambda_c \to J/\psi \Lambda$, $\bar{D}_s\Lambda_c \to \bar{D}\Xi_c$, and $\bar{D}_s\Lambda_c \to \bar{D}_s\Lambda_c$ scattering processes, respectively. The corresponding scattering matrix $T^{\prime}$ is obtained from   the potential $V^{\prime}$ with Eq.~\eqref{T}.  In addition to the decay vertices of $B^{-} \rightarrow J / \psi \Lambda \bar{p}$ and $B^{-} \rightarrow  \bar{D}\Xi_c \bar{p}$ considered in scenario I, the vertex of the decay $B^{-} \rightarrow  \bar{D}_s \Lambda_c \bar{p}$ is also taken into account in scenario II, which is characterized by a constant $e$.   The final-state $\bar{D}_s \Lambda_c$ of the decay $B^{-} \rightarrow  \bar{D}_s \Lambda_c \bar{p}$ can scatter into the final state $J/\psi \Lambda$, described by  $eG_{\bar{D_{s}}\Lambda_{c}}T^{\prime}_{33}$. The amplitude $T_{B}^{\prime}$ in scenario II  is given by  
\begin{equation}
    T_{B}^{\prime}=d+dG_{J/\psi \Lambda}T_{11}^{\prime}+fG_{\bar{D}\Xi_{c}}T_{21}^{\prime}+eG_{\bar{D_{s}}\Lambda_{c}}T_{31}^{\prime}.  
    \label{T2}
\end{equation}

\begin{table}[!h]
\setlength{\tabcolsep}{10pt}
\centering
\caption{Values of the potential parameters for scenario I (Unit:GeV$^{-1}$). 
\label{CDDbs}}
\begin{tabular}{ccccccc}
  \hline\hline
    $d$  &$f$ & $a$  & $b$  & $c$  &$x_{back}$\\
  \hline
     $20$ & $-19$ & $-3.9$ & $-19.4$ & $41.4$ &$25$\\
 \hline \hline
\end{tabular}
\end{table}

\begin{figure}
\includegraphics[width=0.6\textwidth]{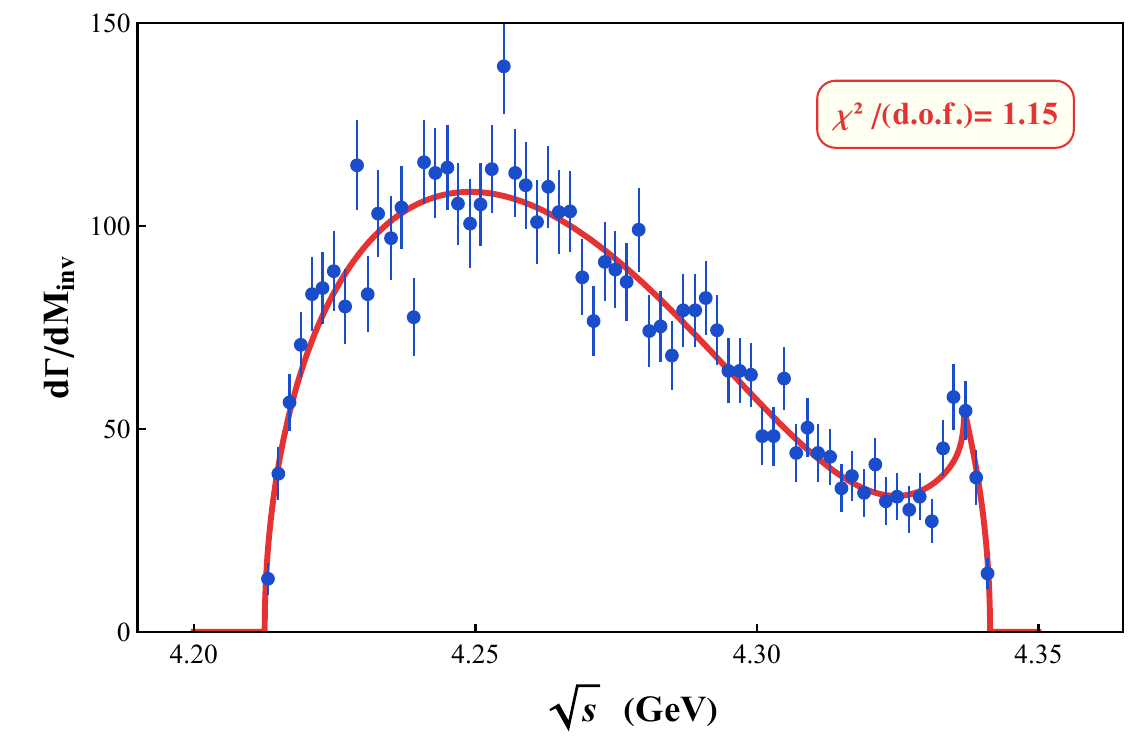} 
\caption{$J/\psi \Lambda$ invariant mass spectrum for Scenario I.}
\label{back1}
\end{figure}

As shown in Fig.~\ref{back1}, the $J/\psi \Lambda$ mass spectrum is reproduced in scenario I, yielding $\chi^{2}/({\rm d.o.f.}) = 1.15$. The corresponding fit parameters are listed in Table~\ref{CDDbs}. Our results indicate that the off-diagonal potential $\bar{D}\Xi_c \to J/\psi \Lambda$ ($c = 41.4$~GeV$^{-1}$) is significantly more attractive than both diagonal potentials: $J/\psi\Lambda \to J/\psi\Lambda$ ($a = -3.9$~GeV$^{-1}$) and $\bar{D}\Xi_c \to \bar{D}\Xi_c$ ( $b = -19.4$~GeV$^{-1}$). It is difficult to interpret this behavior, since off-diagonal potentials typically play a less important role in resonant-state formation compared to diagonal ones. Consequently, we conclude that a satisfactory description cannot be achieved in scenario I.

\begin{figure}
    \centering
    \includegraphics[width=0.6\textwidth]{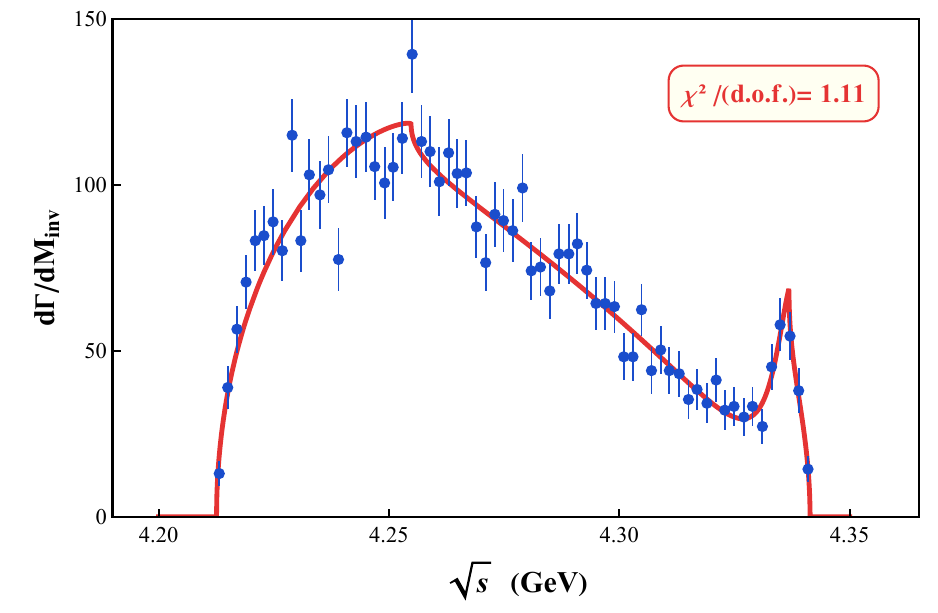} 
    \caption{ $J/\psi \Lambda$ invariant mass spectrum for Scenario II.} 
    \label{fig:three_coupled}
\end{figure}

In scenario II, due to the small background contribution, we simplify the model by reducing the number of parameters and neglecting the background to avoid overfitting. The fitting results are presented in Fig.~\ref{fig:three_coupled} with $\chi^{2}/({\rm d.o.f.}) = 1.11$, smaller than that in scenario I, indicating that the results in scenario II are only marginally better than those in scenario I.    The values of the corresponding parameters are given in Table~\ref{table2}. In scenario II, the off-diagonal potentials are smaller in magnitude than the diagonal ones, which is more physically plausible. 

\begin{table}[!h]
\setlength{\tabcolsep}{10pt}
\centering
\caption{Values of the potential parameters for scenario II (Unit: GeV$^{-1}$). 
\label{table2}}
\begin{tabular}{ccccccccc}
  \hline\hline
   $d$  &$f$ &$e $ & $a$  & $b$  & $c$  & $x$  & $y$   & $z$\\
  \hline
   $-36$ & $5.9$ & $9.3$ & $0$ & $-37$ &$7.1$ &$28.1$ &$-12.1$ &$0$\\
 \hline \hline
\end{tabular}
\end{table}

\begin{figure}
\includegraphics[width=0.6\textwidth]{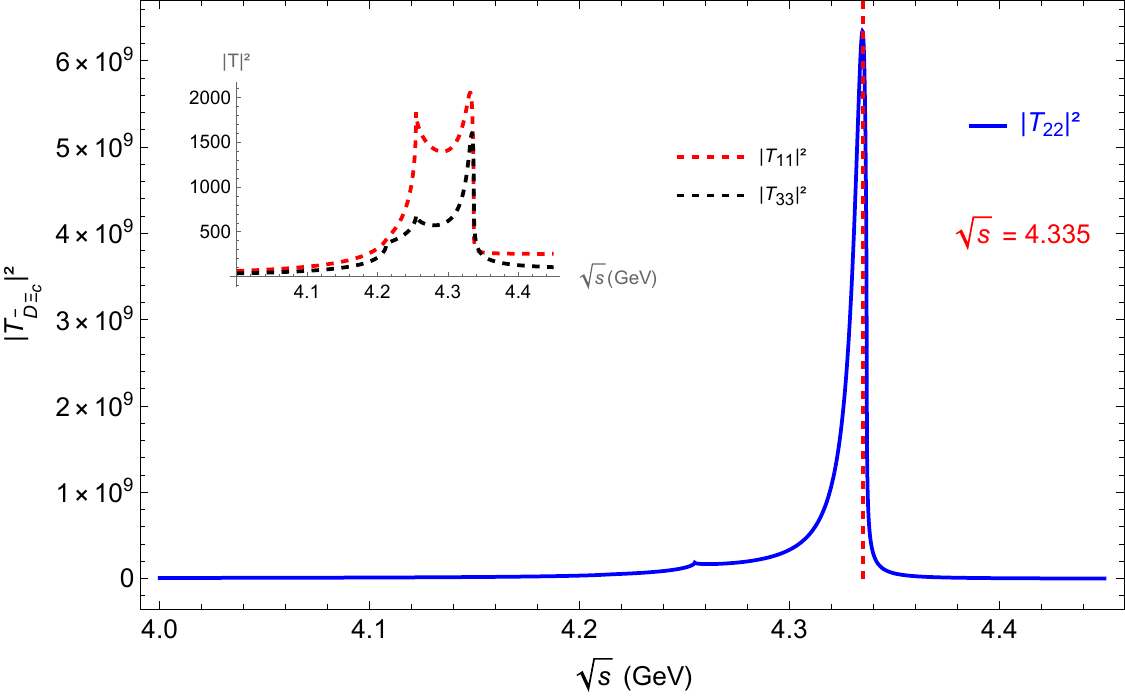} 
\caption{Modulus squared of the $\bar{D}\Xi_c$ $T$-matrix for Scenario II.  }
\label{T22}
\end{figure}

\begin{figure}
\includegraphics[width=0.6\textwidth]{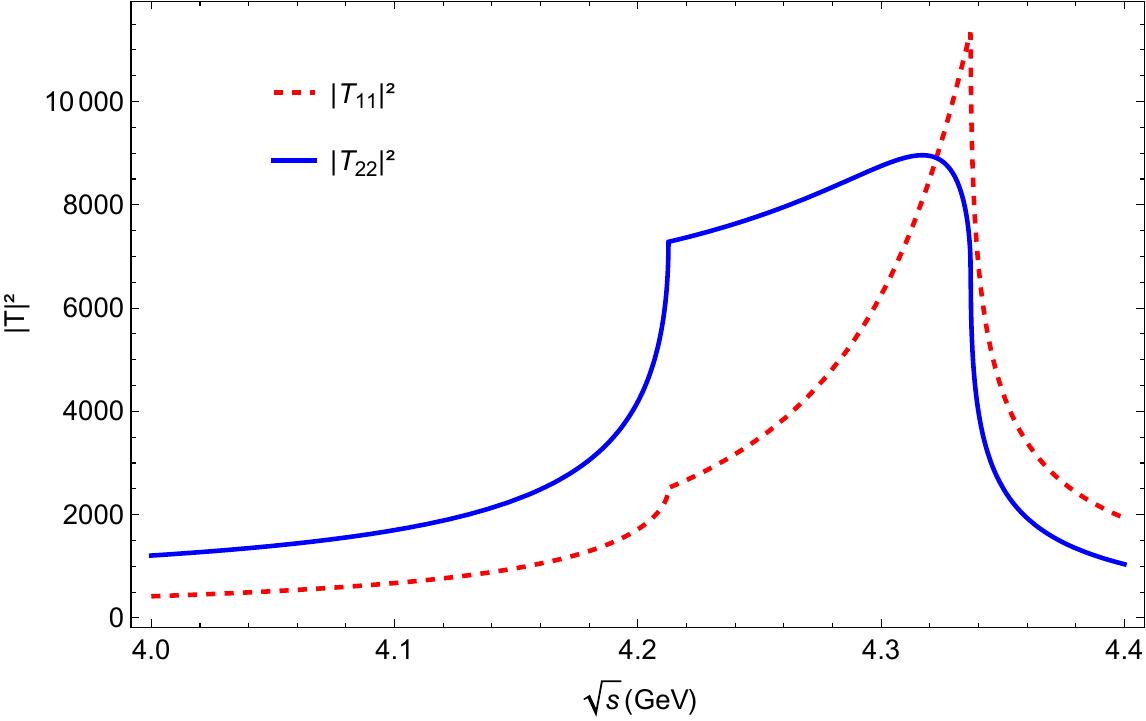} 
\caption{Modulus squared of the  $T$-matrix for Scenario I.  }
\label{fig:T two}
\end{figure}

In the following, we explain why the results in scenario II are more reasonable than those in scenario I by examining other physical observables related to the $T$ matrix.
In Fig.~\ref{T22}, we present the squared diagonal elements of the $T$ matrix.
A distinct peak near $4.335$ GeV is seen in the lineshapes of $\left|T_{J/\psi \Lambda}\right|^2$, $\left|T_{\bar{D} \Xi_c}\right|^2$, and $\left|T_{\bar{D}_s \Lambda_c}\right|^2$. This peak lies below the $\bar{D} \Xi_c$ threshold, indicating the presence of a bound state.  However, this prominent peak disappears entirely when the $\bar{D}_s \Lambda_c$ channel is omitted from the calculation, as shown in Fig.~\ref{fig:T two}, and the lineshapes of $J/\psi\Lambda$ and $\bar{D} \Xi_c$ undergo significant change. The behavior of their lineshapes does not support the existence of a genuine state. The Review of Particle Physics (RPP) reports the mass of the $P_{\psi s}^{\Lambda}(4338)$ as $4338.2 \pm 0.8 \mathrm{MeV}$ and its width as $7.0 \pm$1.8 MeV.   Then, we search for likely poles around the $\bar{D}\Xi_c$ mass threshold in these two scenarios.    In scenario I, the associated pole of the $T$-matrix yields an unreasonably large width of about 100 MeV. In contrast, scenario II provides a much more plausible width of approximately 10 MeV. The pole obtained in scenario II is more consistent with the experimental data than that in scenario I. Our analysis strongly supports that the peak at 4.335 GeV originates from a bound state formed due to the coupled-channel dynamics with $\bar{D}_s \Lambda_c$. This conclusion is consistent with the findings of Refs.~\cite{Zhu:2022wpi,Nakamura:2022gtu}.

\begin{figure}
\includegraphics[width=0.6\textwidth]{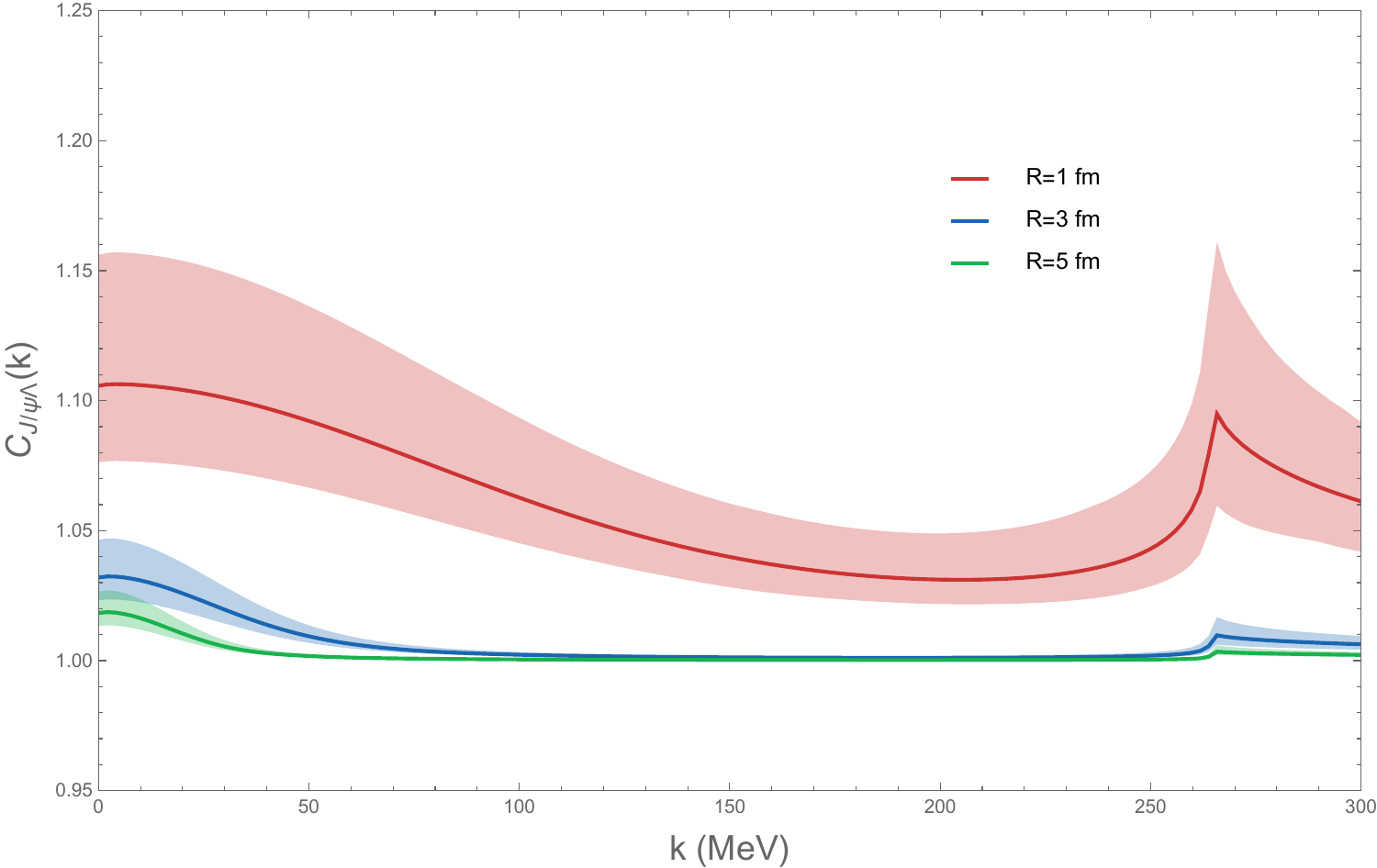} 
\caption{$J/\psi \Lambda$ correlation functions obtained with the parameters of Table~\ref{table2}.}
\label{cf2}
\end{figure}

\begin{figure}
\includegraphics[width=0.6\textwidth]{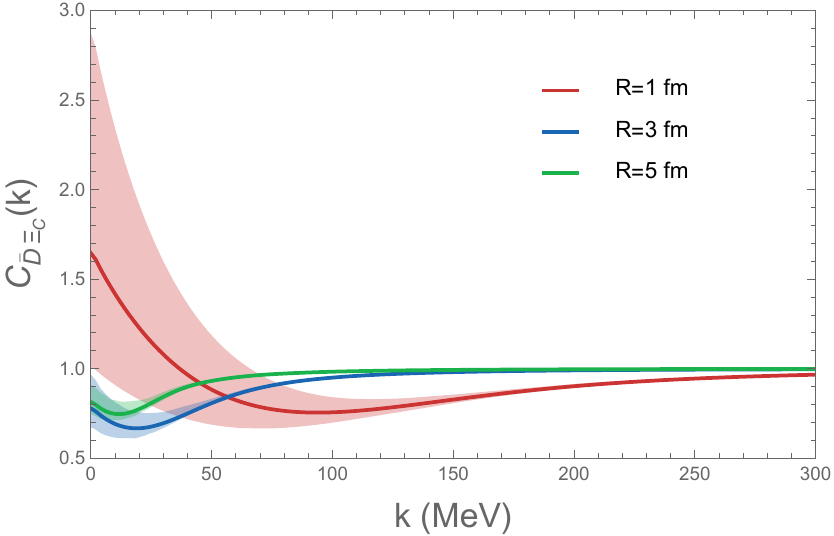} 
\caption{$\bar{D}\Xi_c$ correlation functions obtained with the parameters of Table~\ref{table2}.}
\label{cf3}
\end{figure}

To test the above conclusion in the future, we predict additional physical observables associated with the scattering matrix $T$ in scenario II. First, we compute the CFs using the obtained potential with source sizes ranging from $1$ to $5$ fm. The $J/\psi \Lambda$ CFs are shown in   Fig.~\ref{cf2}, and those of $\bar{D}\Xi_c$ are shown in Fig.~\ref{cf3}, where the peak in Fig.~\ref{cf2} corresponds to the threshold effect of the $\bar{D}_s\Lambda_c$ channel and the error band comes from the 1-$\sigma$ uncertainty of each parameter in the nonlinear fitting.  We can see that the $J/\psi \Lambda$ interaction is weakly attractive (more than unity) and the $\bar{D}\Xi_c$ interaction is strongly attractive (less than unity), whose lineshapes are consistent with the features expected from square well potentials~\cite{Liu:2023uly}. Then, we calculate the scattering lengths using the scattering matrix $T$: $\mathrm{Re}[a_{J/\psi \Lambda}]= -0.17$ fm, $\mathrm{Re}[a_{\bar{D}\Xi_c}]= 1.22$ fm, and $\mathrm{Re}[a_{\bar{D}_s\Lambda_c}]= -0.19$ fm. In our convention, a positive scattering length corresponds to a repulsive or attractive interaction capable of generating a bound state, and a negative value indicates a purely attractive interaction~\cite{Macedo-Lima:2023fzp}. The obtained $\bar{D}\Xi_c$ scattering length suggests the existence of a bound state near the $\bar{D}\Xi_c$ threshold as well.

The $\bar{D}\Xi_c$ and $\bar{D}_s\Lambda_c$ systems are related  to the $\bar{D}\Lambda_c$ system through SU(3)-flavor symmetry. In the following, we investigate the potentials of the coupled-channel $\bar{D}\Lambda_c$–$J/\psi N$ system, where the strengths of these potentials are determined by fitting to the $J/\psi p$ mass spectrum of the decay $\Lambda_b \to J/\psi p K$ in energy region $\sqrt{s} < 4.30\ \text{GeV}$~\cite{LHCb:2019kea}.  In this region, there exist three coupled channels: $J/\psi N-\bar{D}\Lambda_c-\bar{D}^{*}\Lambda_c$. Following Ref.~\cite{Pan:2023hrk}, the contact-range potentials of the $J/\psi N-\bar{D}\Lambda_c-\bar{D}^{*}\Lambda_c$ system  are written as
\begin{equation}
\mathbf{V}_{J/\psi N-\bar{D}\Lambda_c-\bar{D}^{*}\Lambda_c}=\left(\begin{array}{ccc}
0 & 0 & \frac{1}{2}g \\
&b^{\prime} & \frac{\sqrt{3}}{2}g \\
& & b^{\prime}
\end{array}\right).
\label{V3}
\end{equation}

\begin{figure}
\includegraphics[width=0.6\textwidth]{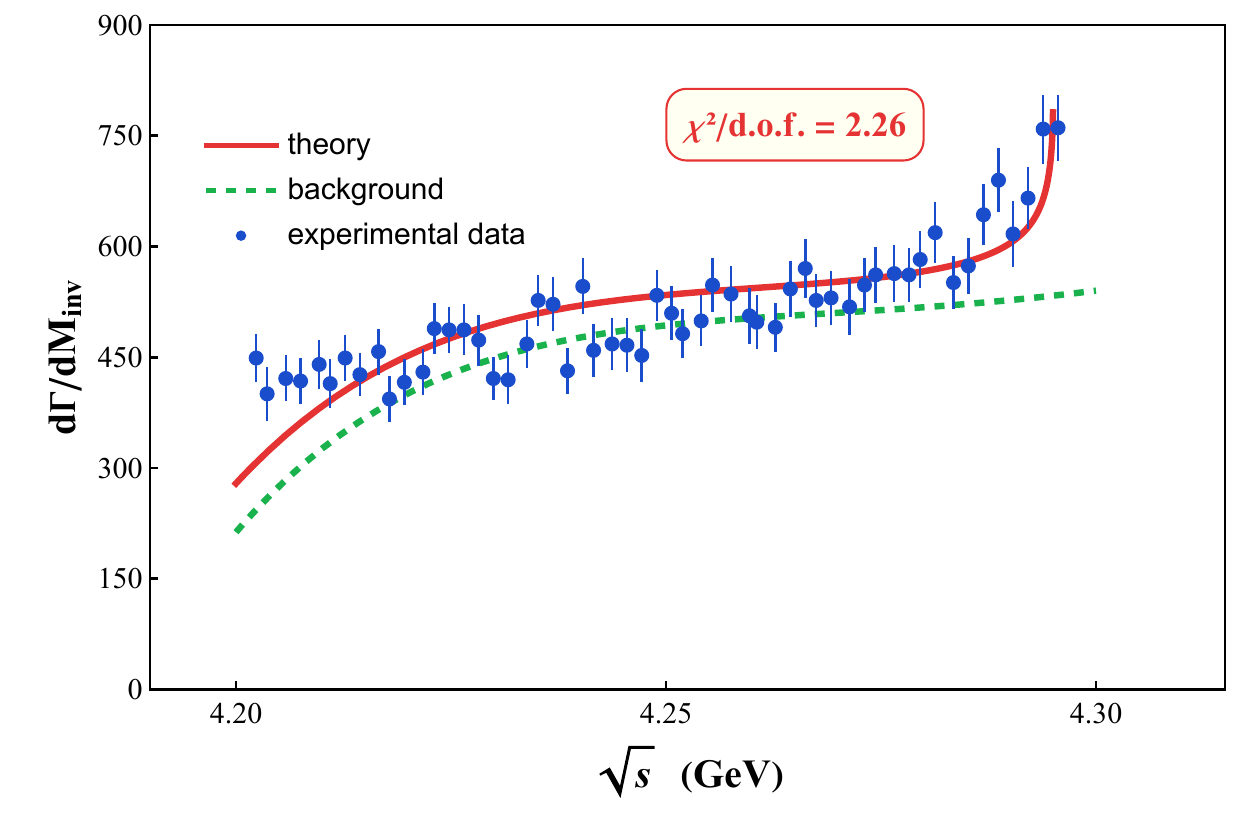} 
\caption{ $J/\psi N$ invariant mass spectrum.}
\label{JN}
\end{figure}

The amplitude for the decay $\Lambda_b \to J/\psi p K$, including $J/\psi p$ rescattering effects, has a form similar to that given in Eq.~\eqref{T2}.
For this process, the background is approximated by a sixth-order polynomial, following Ref.~\cite{LHCb:2019kea}.
Our fit results are presented in Fig.~\ref{JN}, yielding $\chi^{2}/(\mathrm{d.o.f.})=2.26$.
The two unknown couplings in Eq.~\eqref{V3} are determined as $b^{\prime}= -31\ \mathrm{GeV}^{-1}$ and $g=-21\ \mathrm{GeV}^{-1}$.
We further compute the $J/\psi N$ scattering length as $a_{J/\psi N} = -0.03\ \mathrm{fm}$, indicating a very weak attractive interaction, consistent with the coupled-channel results of Ref.~\cite{Wu:2025rur}.
Compared with the scattering length $\mathrm{Re}[a_{J/\psi \Lambda}]= -0.17\ \mathrm{fm}$, the $J/\psi\Lambda$ potential is found to be more attractive than the $J/\psi p$ potential.
Additionally, we obtain the scattering lengths $a_{\bar{D}^{*}\Lambda_c}=-1.39\ \mathrm{fm}$ and $a_{\bar{D}\Lambda_c}=-1.89\ \mathrm{fm}$, both suggesting weakly attractive interactions.
Hence, no bound state exists near the $\bar{D}^{(*)}\Lambda_c$ mass thresholds, consistent with Refs.~\cite{Xiao:2020frg,Yalikun:2021bfm,Pan:2023hrk}. Similarly, the $\bar{D}\Xi_c$ potential is more attractive than the  $\bar{D}\Lambda_c$ potential.

\section{Summary and outlook}

Hadron-hadron interactions play a significant role in understanding the spectroscopy of exotic states discovered experimentally in recent years. The invariant mass spectrum of a hadron pair from inclusive or exclusive decays provides essential input for determining such interactions.   In this work, we focused on the   $J/\psi \Lambda$ invariant mass spectrum of the exclusive decay $B^{-} \rightarrow J/\psi \Lambda \bar{p}$, which can be factorized into short-range and long-range contributions mediated by weak and strong interactions, respectively. The short-range part is parameterized by an unknown parameter, while the long-range part is described by a contact-range EFT potential that satisfies unitarity, expressed in terms of the $T$-matrix. In our calculations, we consider two scenarios: scenario I (two-channel $J/\psi \Lambda - \bar{D} \Xi_c$)   and scenario II (three-channel  $J/\psi \Lambda - \bar{D} \Xi_c - \bar{D}_s \Lambda_c$). The fit results favor Scenario II over Scenario I. In Scenario II, we obtained a molecular state below the $\bar{D}\Xi_c$ mass threshold,  with a pole position at $4335-5i $ MeV, consistent with the experimentally observed $P_{cs}(4338)$ state.

Utilizing the potentials determined in scenario II, we calculated the scattering lengths: $\mathrm{Re}[a_{J/\psi \Lambda}]= -0.17$ fm, $\mathrm{Re}[a_{\bar{D}\Xi_c}]= 1.22$ fm, and $\mathrm{Re}[a_{\bar{D}_s\Lambda_c}]= -0.19$ fm. The sizable positive   $\bar{D}\Xi_c$ scattering length also suggests the presence of a bound state near the $\bar{D}\Xi_c$ threshold. Furthermore, based on the potentials from scenario II, we predicted the CFs of  $J/\psi \Lambda$  and $ \bar{D} \Xi_c$ for different Gaussian source sizes.  These predictions provide a theoretical reference for future experimental measurements. In particular, the CFs can help to test the coupled-channel potentials and shed light on the dynamics of the pentaquark state $P_{cs}(4338)$.  Finally,  we extended the study to the SU(3)-flavor partner system of  $J/\psi \Lambda - \bar{D} \Xi_c - \bar{D}_s \Lambda_c$.  Using a similar approach,  the potentials of the coupled-channel system $J/\psi N-\bar{D}\Lambda_c-\bar{D}^{*}\Lambda_c$ were determined by fitting the $J/\psi p$ mass spectrum of the decay $\Lambda_b \to J/\psi p K$. The resulting scattering lengths are: $a_{J/\psi N} = -0.03\ \mathrm{fm}$, $a_{\bar{D}^{*}\Lambda_c}=-1.39\ \mathrm{fm}$, and $a_{\bar{D}\Lambda_c}=-1.89\ \mathrm{fm}$, indicating interactions that are less attractive compared to their SU(3)-flavor counterparts.

\section{Acknowledgments}
This work is partly supported by the National Science Foundation of China under Grant No. W2543006 and Nos. 12435007 and 1252200936, and the National Key R\&D Program of China under Grant No. 2023YFA1606703. Ming-Zhu Liu acknowledges support from the National Natural Science Foundation of China under Grant No.12575086. Zhi-Wei Liu acknowledges support from the National Natural Science Foundation of China under Grant No.12405133, No.12347180, China Postdoctoral Science Foundation under Grant No.2023M740189, and the Postdoctoral Fellowship Program of CPSF under Grant No.GZC20233381. Rui-Xiang Shi acknowledges support from the National Natural Science Foundation of China under Grant No. 12405091 and the Natural Science Foundation of Guangxi province under Grant No. 2025GXNSFBA069314.
W.H. Liang and C.W. Xiao acknowledge support from the National Natural Science Foundation of China under Grant Nos. 12575081 and 12365019, and the Natural Science Foundation of Guangxi province under Grant No. 2023JJA110076.

\bibliography{reference}

@article{Nakamura:2022gtu,
    author = "Nakamura, S. X. and Wu, J. -J.",
    title = "{Pole determination of P{\ensuremath{\psi}}s{\ensuremath{\Lambda}}(4338) and possible P{\ensuremath{\psi}}s{\ensuremath{\Lambda}}(4255) in B-{\textrightarrow}J/{\ensuremath{\psi}}{\ensuremath{\Lambda}}p{\textasciimacron}}",
    eprint = "2208.11995",
    archivePrefix = "arXiv",
    primaryClass = "hep-ph",
    doi = "10.1103/PhysRevD.108.L011501",
    journal = "Phys. Rev. D",
    volume = "108",
    number = "1",
    pages = "L011501",
    year = "2023"
}

@article{Pan:2025vyg,
    author = "Pan, Ya-Wen and Liu, Ming-Zhu and Geng, Li-Sheng",
    title = "{Probing the three-body force in hadronic systems with specific charge parity}",
    eprint = "2512.01468",
    archivePrefix = "arXiv",
    primaryClass = "nucl-th",
    month = "12",
    year = "2025"
}

@article{Zhu:2022wpi,
    author = "Zhu, Jun-Tao and Kong, Shu-Yi and He, Jun",
    title = "{P{\ensuremath{\psi}}s{\ensuremath{\Lambda}}(4459) and P{\ensuremath{\psi}}s{\ensuremath{\Lambda}}(4338) as molecular states in J/{\ensuremath{\psi}}{\ensuremath{\Lambda}} invariant mass spectra}",
    eprint = "2211.06232",
    archivePrefix = "arXiv",
    primaryClass = "hep-ph",
    doi = "10.1103/PhysRevD.107.034029",
    journal = "Phys. Rev. D",
    volume = "107",
    number = "3",
    pages = "034029",
    year = "2023"
}

@article{Sibirtsev:2005ex,
    author = "Sibirtsev, A. and Voloshin, M. B.",
    title = "{The Interaction of slow J/psi and psi' with nucleons}",
    eprint = "hep-ph/0502068",
    archivePrefix = "arXiv",
    reportNumber = "FZJ-IKP-TH-2005-6, FTPI-MINN-05-03, UMN-TH-2344-05",
    doi = "10.1103/PhysRevD.71.076005",
    journal = "Phys. Rev. D",
    volume = "71",
    pages = "076005",
    year = "2005"
}

@article{Hayashigaki:1998ey,
    author = "Hayashigaki, Arata",
    title = "{J / psi nucleon scattering length and in-medium mass shift of J / psi in QCD sum rule analysis}",
    eprint = "nucl-th/9811092",
    archivePrefix = "arXiv",
    doi = "10.1143/PTP.101.923",
    journal = "Prog. Theor. Phys.",
    volume = "101",
    pages = "923--935",
    year = "1999"
}

@article{Pentchev:2020kao,
    author = "Pentchev, Lubomir and Strakovsky, Igor I.",
    title = "{$J/\psi$-$p$ Scattering Length from the Total and Differential Photoproduction Cross Sections}",
    eprint = "2009.04502",
    archivePrefix = "arXiv",
    primaryClass = "hep-ph",
    doi = "10.1140/epja/s10050-021-00364-4",
    journal = "Eur. Phys. J. A",
    volume = "57",
    number = "2",
    pages = "56",
    year = "2021"
}

@article{Macedo-Lima:2023fzp,
    author = "Mac{\^e}do-Lima, Mathias and Madeira, Lucas",
    title = "{Scattering length and effective range of microscopic two-body potentials}",
    eprint = "2303.04591",
    archivePrefix = "arXiv",
    primaryClass = "quant-ph",
    doi = "10.1590/1806-9126-RBEF-2023-0079",
    journal = "Rev. Bras. Ens. Fis.",
    volume = "45",
    pages = "e20230079",
    year = "2023"
}

@article{Lyu:2024ttm,
    author = "Lyu, Yan and Doi, Takumi and Hatsuda, Tetsuo and Sugiura, Takuya",
    title = "{Nucleon-charmonium interactions from lattice QCD}",
    eprint = "2410.22755",
    archivePrefix = "arXiv",
    primaryClass = "hep-lat",
    reportNumber = "RIKEN-iTHEMS-Report-24",
    doi = "10.1016/j.physletb.2024.139178",
    journal = "Phys. Lett. B",
    volume = "860",
    pages = "139178",
    year = "2025"
}

@article{Wang:2022xpw,
    author = "Wang, Xiao-Yun and Zeng, Fancong and Strakovsky, Igor I.",
    title = "{{\ensuremath{\psi}}(*)p scattering length based on near-threshold charmonium photoproduction}",
    eprint = "2205.07661",
    archivePrefix = "arXiv",
    primaryClass = "hep-ph",
    doi = "10.1103/PhysRevC.106.015202",
    journal = "Phys. Rev. C",
    volume = "106",
    number = "1",
    pages = "015202",
    year = "2022"
}

@article{Gryniuk:2016mpk,
    author = "Gryniuk, Oleksii and Vanderhaeghen, Marc",
    title = "{Accessing the real part of the forward $J/\psi$-p scattering amplitude from $J/\psi$ photoproduction on protons around threshold}",
    eprint = "1608.08205",
    archivePrefix = "arXiv",
    primaryClass = "hep-ph",
    doi = "10.1103/PhysRevD.94.074001",
    journal = "Phys. Rev. D",
    volume = "94",
    number = "7",
    pages = "074001",
    year = "2016"
}

@article{Kaidalov:1992hd,
    author = "Kaidalov, A. B. and Volkovitsky, P. E.",
    title = "{Heavy quarkonia interactions with nucleons and nuclei}",
    doi = "10.1103/PhysRevLett.69.3155",
    journal = "Phys. Rev. Lett.",
    volume = "69",
    pages = "3155--3156",
    year = "1992"
}

@article{Wu:2024xwy,
    author = "Wu, Bing and Dong, Xiang-Kun and Du, Meng-Lin and Guo, Feng-Kun and Zou, Bing-Song",
    title = "{Deciphering the mechanism of $J/ψ$-nucleon scattering}",
    eprint = "2410.19526",
    archivePrefix = "arXiv",
    primaryClass = "hep-ph",
    doi = "10.1016/j.fmre.2025.07.005",
    month = "10",
    year = "2024"
}

@article{Du:2020bqj,
    author = "Du, Meng-Lin and Baru, Vadim and Guo, Feng-Kun and Hanhart, Christoph and Mei{\ss}ner, Ulf-G and Nefediev, Alexey and Strakovsky, Igor",
    title = "{Deciphering the mechanism of near-threshold $J/\psi$ photoproduction}",
    eprint = "2009.08345",
    archivePrefix = "arXiv",
    primaryClass = "hep-ph",
    doi = "10.1140/epjc/s10052-020-08620-5",
    journal = "Eur. Phys. J. C",
    volume = "80",
    number = "11",
    pages = "1053",
    year = "2020"
}

@inproceedings{Wu:2025rur,
    author = "Wu, Bing and Dong, Xiang-Kun and Du, Meng-Lin and Guo, Feng-Kun and Zou, Bing-Song",
    title = "{The $J/\psi$-nucleon interaction mechanism: A theoretical study based on scattering length}",
    booktitle = "{11th International Workshop on Chiral Dynamics}",
    eprint = "2503.00702",
    archivePrefix = "arXiv",
    primaryClass = "hep-ph",
    month = "3",
    year = "2025"
}

@article{Shen:2025qpj,
    author = "Shen, Yi-bo and Liu, Zhi-Wei and Lu, Jun-Xu and Liu, Ming-Zhu and Geng, Li-Sheng",
    title = "{Probing the structure of exotic hadrons through correlation functions}",
    eprint = "2506.23476",
    archivePrefix = "arXiv",
    primaryClass = "hep-ph",
    month = "6",
    year = "2025"
}

@article{ALICE:2025aur,
    author = "Acharya, Shreyasi and others",
    collaboration = "ALICE",
    title = "{Investigating the $\mathbf {\text {p--}\uppi ^{\pm }}$ and $\mathbf {\text {p--p--}\uppi ^{\pm }}$ dynamics with femtoscopy in pp collisions at $\mathbf {\sqrt{\textit{s}}=13}$~TeV}",
    eprint = "2502.20200",
    archivePrefix = "arXiv",
    primaryClass = "nucl-ex",
    reportNumber = "CERN-EP-2025-034",
    doi = "10.1140/epja/s10050-025-01615-4",
    journal = "Eur. Phys. J. A",
    volume = "61",
    number = "8",
    pages = "194",
    year = "2025"
}

@article{Kievsky:2023maf,
    author = "Kievsky, A. and Garrido, E. and Viviani, M. and Marcucci, L. E. and Serksnyte, L. and Del Grande, R.",
    title = "{nnn and ppp correlation functions}",
    eprint = "2310.10428",
    archivePrefix = "arXiv",
    primaryClass = "nucl-th",
    doi = "10.1103/PhysRevC.109.034006",
    journal = "Phys. Rev. C",
    volume = "109",
    number = "3",
    pages = "034006",
    year = "2024"
}

@article{Burns:2022uha,
    author = "Burns, T. J. and Swanson, E. S.",
    title = "{The LHCb state P{\ensuremath{\psi}}s{\ensuremath{\Lambda}}(4338) as a triangle singularity}",
    eprint = "2208.05106",
    archivePrefix = "arXiv",
    primaryClass = "hep-ph",
    doi = "10.1016/j.physletb.2023.137715",
    journal = "Phys. Lett. B",
    volume = "838",
    pages = "137715",
    year = "2023"
}

@article{Liu:2025oar,
    author = "Liu, Zhi-Wei and Ge, Duo-Lun and Lu, Jun-Xu and Liu, Ming-Zhu and Geng, Li-Sheng",
    title = "{Charmonium-nucleon femtoscopic correlation function}",
    eprint = "2504.04853",
    archivePrefix = "arXiv",
    primaryClass = "hep-ph",
    doi = "10.1103/3bdh-blwh",
    journal = "Phys. Rev. D",
    volume = "112",
    number = "5",
    pages = "054019",
    year = "2025"
}

@article{Liu:2023wfo,
    author = "Liu, Zhi-Wei and Lu, Jun-Xu and Liu, Ming-Zhu and Geng, Li-Sheng",
    title = "{Distinguishing the spins of Pc(4440) and Pc(4457) with femtoscopic correlation functions}",
    eprint = "2305.19048",
    archivePrefix = "arXiv",
    primaryClass = "hep-ph",
    doi = "10.1103/PhysRevD.108.L031503",
    journal = "Phys. Rev. D",
    volume = "108",
    number = "3",
    pages = "L031503",
    year = "2023"
}

@article{Wang:2025sic,
    author = "Wang, Zhi-Gang",
    title = "{Review of the QCD sum rules for exotic states}",
    eprint = "2502.11351",
    archivePrefix = "arXiv",
    primaryClass = "hep-ph",
    month = "2",
    year = "2025"
}

@article{Liu:2024uxn,
    author = "Liu, Ming-Zhu and Pan, Ya-Wen and Liu, Zhi-Wei and Wu, Tian-Wei and Lu, Jun-Xu and Geng, Li-Sheng",
    title = "{Three ways to decipher the nature of exotic hadrons: Multiplets, three-body hadronic molecules, and correlation functions}",
    eprint = "2404.06399",
    archivePrefix = "arXiv",
    primaryClass = "hep-ph",
    doi = "10.1016/j.physrep.2024.12.001",
    journal = "Phys. Rept.",
    volume = "1108",
    pages = "1--108",
    year = "2025"
}

@article{Molina:2023oeu,
    author = "Molina, R. and Liu, Zhi-Wei and Geng, Li-Sheng and Oset, E.",
    title = "{Correlation function for the $a_0(980)$}",
    eprint = "2312.11993",
    archivePrefix = "arXiv",
    primaryClass = "hep-ph",
    doi = "10.1140/epjc/s10052-024-12694-w",
    journal = "Eur. Phys. J. C",
    volume = "84",
    number = "3",
    pages = "328",
    year = "2024"
}

@article{Eichten:1974af,
    author = "Eichten, E. and Gottfried, K. and Kinoshita, T. and Kogut, John B. and Lane, K. D. and Yan, Tung-Mow",
    title = "{The Spectrum of Charmonium}",
    reportNumber = "PRINT-74-1715 (CORNELL)",
    doi = "10.1103/PhysRevLett.34.369",
    journal = "Phys. Rev. Lett.",
    volume = "34",
    pages = "369--372",
    year = "1975",
    note = "[Erratum: Phys.Rev.Lett. 36, 1276 (1976)]"
}

@article{Xie:2025xew,
    author = "Xie, Jia-Ming and Liu, Zhi-Wei and Lu, Jun-Xu and Liang, Haozhao and Molina, Raquel and Geng, Li-Sheng",
    title = "{Chiral Evolution and Femtoscopic Signatures of the $K_1(1270)$ Resonance}",
    eprint = "2511.14380",
    archivePrefix = "arXiv",
    primaryClass = "hep-ph",
    month = "11",
    year = "2025"
}

@article{Liu:2025wwx,
    author = "Liu, Hao-Nan and Liu, Zhi-Wei and Abreu, Luciano and Geng, Li-Sheng",
    title = "{Traces of the $X(3960)$ state in the femtoscopic $D_s^+ D_s^- $ correlations}",
    eprint = "2511.19098",
    archivePrefix = "arXiv",
    primaryClass = "hep-ph",
    month = "11",
    year = "2025"
}

@article{Kamiya:2019uiw,
    author = "Kamiya, Yuki and Hyodo, Tetsuo and Morita, Kenji and Ohnishi, Akira and Weise, Wolfram",
    title = "{$K^-p$ Correlation Function from High-Energy Nuclear Collisions and Chiral SU(3) Dynamics}",
    eprint = "1911.01041",
    archivePrefix = "arXiv",
    primaryClass = "nucl-th",
    doi = "10.1103/PhysRevLett.124.132501",
    journal = "Phys. Rev. Lett.",
    volume = "124",
    number = "13",
    pages = "132501",
    year = "2020"
}

@article{Liu:2023uly,
    author = "Liu, Zhi-Wei and Lu, Jun-Xu and Geng, Li-Sheng",
    title = "{Study of the DK interaction with femtoscopic correlation functions}",
    eprint = "2302.01046",
    archivePrefix = "arXiv",
    primaryClass = "hep-ph",
    doi = "10.1103/PhysRevD.107.074019",
    journal = "Phys. Rev. D",
    volume = "107",
    number = "7",
    pages = "074019",
    year = "2023"
}

@article{Oset:2016lyh,
    author = "Oset, Eulogio and others",
    title = "{Weak decays of heavy hadrons into dynamically generated resonances}",
    eprint = "1601.03972",
    archivePrefix = "arXiv",
    primaryClass = "hep-ph",
    doi = "10.1142/S0218301316300010",
    journal = "Int. J. Mod. Phys. E",
    volume = "25",
    pages = "1630001",
    year = "2016"
}

@article{Xiao:2020frg,
    author = "Xiao, C. W. and Lu, J. X. and Wu, J. J. and Geng, L. S.",
    title = "{How to reveal the nature of three or more pentaquark states}",
    eprint = "2007.12106",
    archivePrefix = "arXiv",
    primaryClass = "hep-ph",
    doi = "10.1103/PhysRevD.102.056018",
    journal = "Phys. Rev. D",
    volume = "102",
    number = "5",
    pages = "056018",
    year = "2020"
}

@article{Lebed:2016hpi,
    author = "Lebed, Richard F. and Mitchell, Ryan E. and Swanson, Eric S.",
    title = "{Heavy-Quark QCD Exotica}",
    eprint = "1610.04528",
    archivePrefix = "arXiv",
    primaryClass = "hep-ph",
    doi = "10.1016/j.ppnp.2016.11.003",
    journal = "Prog. Part. Nucl. Phys.",
    volume = "93",
    pages = "143--194",
    year = "2017"
}

@article{Ali:2017jda,
    author = {Ali, Ahmed and Lange, Jens S\"oren and Stone, Sheldon},
    title = "{Exotics: Heavy Pentaquarks and Tetraquarks}",
    eprint = "1706.00610",
    archivePrefix = "arXiv",
    primaryClass = "hep-ph",
    reportNumber = "DESY-17-071",
    doi = "10.1016/j.ppnp.2017.08.003",
    journal = "Prog. Part. Nucl. Phys.",
    volume = "97",
    pages = "123--198",
    year = "2017"
}

@article{Olsen:2017bmm,
    author = "Olsen, Stephen Lars and Skwarnicki, Tomasz and Zieminska, Daria",
    title = "{Nonstandard heavy mesons and baryons: Experimental evidence}",
    eprint = "1708.04012",
    archivePrefix = "arXiv",
    primaryClass = "hep-ph",
    doi = "10.1103/RevModPhys.90.015003",
    journal = "Rev. Mod. Phys.",
    volume = "90",
    number = "1",
    pages = "015003",
    year = "2018"
}

@article{Guo:2017jvc,
    author = "Guo, Feng-Kun and Hanhart, Christoph and Mei\ss{}ner, Ulf-G. and Wang, Qian and Zhao, Qiang and Zou, Bing-Song",
    title = "{Hadronic molecules}",
    eprint = "1705.00141",
    archivePrefix = "arXiv",
    primaryClass = "hep-ph",
    doi = "10.1103/RevModPhys.90.015004",
    journal = "Rev. Mod. Phys.",
    volume = "90",
    number = "1",
    pages = "015004",
    year = "2018",
    note = "[Erratum: Rev.Mod.Phys. 94, 029901 (2022)]"
}

@article{Brambilla:2019esw,
    author = "Brambilla, Nora and Eidelman, Simon and Hanhart, Christoph and Nefediev, Alexey and Shen, Cheng-Ping and Thomas, Christopher E. and Vairo, Antonio and Yuan, Chang-Zheng",
    title = "{The $XYZ$ states: experimental and theoretical status and perspectives}",
    eprint = "1907.07583",
    archivePrefix = "arXiv",
    primaryClass = "hep-ex",
    reportNumber = "TUM-EFT 125/19",
    doi = "10.1016/j.physrep.2020.05.001",
    journal = "Phys. Rept.",
    volume = "873",
    pages = "1--154",
    year = "2020"
}

@article{Chen:2016qju,
    author = "Chen, Hua-Xing and Chen, Wei and Liu, Xiang and Zhu, Shi-Lin",
    title = "{The hidden-charm pentaquark and tetraquark states}",
    eprint = "1601.02092",
    archivePrefix = "arXiv",
    primaryClass = "hep-ph",
    doi = "10.1016/j.physrep.2016.05.004",
    journal = "Phys. Rept.",
    volume = "639",
    pages = "1--121",
    year = "2016"
}

@article{LHCb:2019kea,
    author = "Aaij, Roel and others",
    collaboration = "LHCb",
    title = "{Observation of a narrow pentaquark state, $P_c(4312)^+$, and of two-peak structure of the $P_c(4450)^+$}",
    eprint = "1904.03947",
    archivePrefix = "arXiv",
    primaryClass = "hep-ex",
    reportNumber = "LHCb-PAPER-2019-014 CERN-EP-2019-058",
    doi = "10.1103/PhysRevLett.122.222001",
    journal = "Phys. Rev. Lett.",
    volume = "122",
    number = "22",
    pages = "222001",
    year = "2019"
}

@article{Guo:2019twa,
    author = "Guo, Feng-Kun and Liu, Xiao-Hai and Sakai, Shuntaro",
    archivePrefix = "arXiv",
    doi = "10.1016/j.ppnp.2020.103757",
    eprint = "1912.07030",
    journal = "Prog. Part. Nucl. Phys.",
    pages = "103757",
    primaryClass = "hep-ph",
    title = "{Threshold cusps and triangle singularities in hadronic reactions}",
    volume = "112",
    year = "2020"
}

@article{Ikeno:2023ojl,
    author = "Ikeno, Natsumi and Toledo, Genaro and Oset, Eulogio",
    title = "{Model independent analysis of femtoscopic correlation functions: An application to the Ds0\textasteriskcentered{}(2317)}",
    eprint = "2305.16431",
    archivePrefix = "arXiv",
    primaryClass = "hep-ph",
    doi = "10.1016/j.physletb.2023.138281",
    journal = "Phys. Lett. B",
    volume = "847",
    pages = "138281",
    year = "2023"
}

@article{Gell-Mann:1964ewy,
    author = "Gell-Mann, Murray",
    title = "{A Schematic Model of Baryons and Mesons}",
    doi = "10.1016/S0031-9163(64)92001-3",
    journal = "Phys. Lett.",
    volume = "8",
    pages = "214--215",
    year = "1964"
}

@article{Brambilla:2010cs,
    author = "Brambilla, N. and others",
    title = "{Heavy Quarkonium: Progress, Puzzles, and Opportunities}",
    eprint = "1010.5827",
    archivePrefix = "arXiv",
    primaryClass = "hep-ph",
    reportNumber = "SLAC-R-996, TUM-EFT-11-10, CLNS-10-2066, ANL-HEP-PR-10-44, ALBERTA-THY-11-10, CP3-10-37, FZJ-IKP-TH-2010-24, INT-PUB-10-059, JLAB-THY-11-1308, FERMILAB-PUB-10-737-T",
    doi = "10.1140/epjc/s10052-010-1534-9",
    journal = "Eur. Phys. J. C",
    volume = "71",
    pages = "1534",
    year = "2011"
}

@article{Pratt:1990zq,
    author = "Pratt, S. and Csorgo, T. and Zimanyi, J.",
    title = "{Detailed predictions for two pion correlations in ultrarelativistic heavy ion collisions}",
    doi = "10.1103/PhysRevC.42.2646",
    journal = "Phys. Rev. C",
    volume = "42",
    pages = "2646--2652",
    year = "1990"
}

@article{Koonin:1977fh,
    author = "Koonin, S. E.",
    title = "{Proton Pictures of High-Energy Nuclear Collisions}",
    doi = "10.1016/0370-2693(77)90340-9",
    journal = "Phys. Lett. B",
    volume = "70",
    pages = "43--47",
    year = "1977"
}

@article{Fabbietti:2020bfg,
    author = "Fabbietti, L. and Mantovani Sarti, V. and Vazquez Doce, O.",
    title = "{Study of the Strong Interaction Among Hadrons with Correlations at the LHC}",
    eprint = "2012.09806",
    archivePrefix = "arXiv",
    primaryClass = "nucl-ex",
    doi = "10.1146/annurev-nucl-102419-034438",
    journal = "Ann. Rev. Nucl. Part. Sci.",
    volume = "71",
    pages = "377--402",
    year = "2021"
}

@article{Dong:2017gaw,
    author = "Dong, Yubing and Faessler, Amand and Lyubovitskij, Valery E.",
    title = "{Description of heavy exotic resonances as molecular states using phenomenological Lagrangians}",
    doi = "10.1016/j.ppnp.2017.01.002",
    journal = "Prog. Part. Nucl. Phys.",
    volume = "94",
    pages = "282--310",
    year = "2017"
}

@article{Esposito:2016noz,
    author = "Esposito, A. and Pilloni, A. and Polosa, A. D.",
    title = "{Multiquark Resonances}",
    eprint = "1611.07920",
    archivePrefix = "arXiv",
    primaryClass = "hep-ph",
    reportNumber = "JLAB-THY-16-2301",
    doi = "10.1016/j.physrep.2016.11.002",
    journal = "Phys. Rept.",
    volume = "668",
    pages = "1--97",
    year = "2017"
}

@article{Karliner:2017qhf,
    author = "Karliner, Marek and Rosner, Jonathan L. and Skwarnicki, Tomasz",
    title = "{Multiquark States}",
    eprint = "1711.10626",
    archivePrefix = "arXiv",
    primaryClass = "hep-ph",
    doi = "10.1146/annurev-nucl-101917-020902",
    journal = "Ann. Rev. Nucl. Part. Sci.",
    volume = "68",
    pages = "17--44",
    year = "2018"
}

@article{LHCb:2022ogu,
    author = "Aaij, R. and others",
    collaboration = "LHCb",
    title = "{Observation of a J/\ensuremath{\psi}\ensuremath{\Lambda} Resonance Consistent with a Strange Pentaquark Candidate in B-\textrightarrow{}J/\ensuremath{\psi}\ensuremath{\Lambda}p\textasciimacron{} Decays}",
    eprint = "2210.10346",
    archivePrefix = "arXiv",
    primaryClass = "hep-ex",
    reportNumber = "CERN-EP-2022-198, LHCb-PAPER-2022-031",
    doi = "10.1103/PhysRevLett.131.031901",
    journal = "Phys. Rev. Lett.",
    volume = "131",
    number = "3",
    pages = "031901",
    year = "2023"
}

@article{Husken:2024rdk,
    author = {H\"usken, Nils and Norella, Elisabetta Spadaro and Polyakov, Ivan},
    title = "{A Brief Guide to Exotic Hadrons}",
    eprint = "2410.06923",
    archivePrefix = "arXiv",
    primaryClass = "hep-ph",
    month = "10",
    year = "2024"
}

@article{Liu:2024nac,
    author = "Liu, Zhi-Wei and Lu, Jun-Xu and Liu, Ming-Zhu and Geng, Li-Sheng",
    title = "{Femtoscopy can tell whether Zc(3900) and Zcs(3985) are resonances, virtual states, or bound states}",
    eprint = "2404.18607",
    archivePrefix = "arXiv",
    primaryClass = "hep-ph",
    doi = "10.1016/j.scib.2025.09.022",
    journal = "Sci. Bull.",
    volume = "70",
    pages = "3515--3521",
    year = "2025"
}

@article{Liu:2022nec,
    author = "Liu, Zhi-Wei and Li, Kai-Wen and Geng, Li-Sheng",
    title = "{Strangeness S = {\ensuremath{-}}2 baryon-baryon interactions and femtoscopic correlation functions in covariant chiral effective field theory*}",
    eprint = "2201.04997",
    archivePrefix = "arXiv",
    primaryClass = "hep-ph",
    doi = "10.1088/1674-1137/ac988a",
    journal = "Chin. Phys. C",
    volume = "47",
    number = "2",
    pages = "024108",
    year = "2023"
}

@article{Pan:2023hrk,
    author = "Pan, Ya-Wen and Liu, Ming-Zhu and Geng, Li-Sheng",
    title = "{Production rates of hidden-charm pentaquark molecules in \ensuremath{\Lambda}b decays}",
    eprint = "2309.12050",
    archivePrefix = "arXiv",
    primaryClass = "hep-ph",
    doi = "10.1103/PhysRevD.108.114022",
    journal = "Phys. Rev. D",
    volume = "108",
    number = "11",
    pages = "114022",
    year = "2023"
}

@article{Yalikun:2021bfm,
    author = "Yalikun, Nijiati and Lin, Yong-Hui and Guo, Feng-Kun and Kamiya, Yuki and Zou, Bing-Song",
    title = "{Coupled-channel effects of the \ensuremath{\Sigma}c(*)D\textasciimacron{}(*)-\ensuremath{\Lambda}c(2595)D\textasciimacron{} system and molecular nature of the Pc pentaquark states from one-boson exchange model}",
    eprint = "2109.03504",
    archivePrefix = "arXiv",
    primaryClass = "hep-ph",
    doi = "10.1103/PhysRevD.104.094039",
    journal = "Phys. Rev. D",
    volume = "104",
    number = "9",
    pages = "094039",
    year = "2021"
}

@article{Doring:2025sgb,
    author = {D{\"o}ring, Michael and Haidenbauer, Johann and Mai, Maxim and Sato, Toru},
    title = "{Dynamical coupled-channel models for hadron dynamics}",
    eprint = "2505.02745",
    archivePrefix = "arXiv",
    primaryClass = "nucl-th",
    reportNumber = "JLAB-THY-25-4298",
    month = "5",
    year = "2025"
}

@article{Shen:2024npc,
    author = "Shen, Yi-Bo and Liu, Ming-Zhu and Liu, Zhi-Wei and Geng, Li-Sheng",
    title = "{Implication of a negative effective range on the DD\textasciimacron{}* interaction and the nature of X(3872)}",
    eprint = "2409.06409",
    archivePrefix = "arXiv",
    primaryClass = "hep-ph",
    doi = "10.1103/PhysRevD.111.034001",
    journal = "Phys. Rev. D",
    volume = "111",
    number = "3",
    pages = "034001",
    year = "2025"
}

@article{LHCb:2015yax,
    author = "Aaij, Roel and others",
    collaboration = "LHCb",
    title = "{Observation of $J/\psi p$ Resonances Consistent with Pentaquark States in $\Lambda_b^0 \to J/\psi K^- p$ Decays}",
    eprint = "1507.03414",
    archivePrefix = "arXiv",
    primaryClass = "hep-ex",
    reportNumber = "CERN-PH-EP-2015-153, LHCB-PAPER-2015-029",
    doi = "10.1103/PhysRevLett.115.072001",
    journal = "Phys. Rev. Lett.",
    volume = "115",
    pages = "072001",
    year = "2015"
}

@article{Dong:2020hxe,
    author = "Dong, Xiang-Kun and Guo, Feng-Kun and Zou, Bing-Song",
    title = "{Explaining the Many Threshold Structures in the Heavy-Quark Hadron Spectrum}",
    eprint = "2011.14517",
    archivePrefix = "arXiv",
    primaryClass = "hep-ph",
    doi = "10.1103/PhysRevLett.126.152001",
    journal = "Phys. Rev. Lett.",
    volume = "126",
    number = "15",
    pages = "152001",
    year = "2021"
}

@article{Liu:2025nze,
    author = "Liu, Zhi-Wei and Xie, Jia-Ming and Lu, Jun-Xu and Geng, Li-Sheng",
    title = "{Probing the di-$J/Ψ$ interaction and the nature of $X(6200)$ with femtoscopic correlation functions}",
    eprint = "2512.10459",
    archivePrefix = "arXiv",
    primaryClass = "hep-ph",
    month = "12",
    year = "2025"
}

@article{Ge:2025put,
    author = "Ge, Duo-Lun and Liu, Zhi-Wei and Lu, Jun-Xu and Geng, Li-Sheng",
    title = "{Deuteron-deuteron interaction and correlation function}",
    eprint = "2502.18872",
    archivePrefix = "arXiv",
    primaryClass = "nucl-th",
    doi = "10.1103/ttrc-qhv5",
    journal = "Phys. Rev. C",
    volume = "112",
    number = "3",
    pages = "034003",
    year = "2025"
}

@article{Liu:2025rci,
    author = "Liu, Zhi-Wei and Lu, Jun-Xu and Geng, Li-Sheng",
    title = "{The femtoscopic technique{\textemdash}an invaluable tool in studies of exotic hadrons}",
    doi = "10.22323/1.465.0044",
    journal = "PoS",
    volume = "QNP2024",
    pages = "044",
    year = "2025"
}

@article{Wu:2025fzx,
    author = "Wu, Tian-Wei and Liu, Ming-Zhu and Geng, Li-Sheng",
    title = "{Implication of the existence of $J^{PC}=0^{--}$$\bar{D}_sDK$ bound state on nature of $D_{s0}^*(2317)$ and new configuration of exotic state}",
    eprint = "2501.11358",
    archivePrefix = "arXiv",
    primaryClass = "hep-ph",
    month = "1",
    year = "2025"
}

\end{document}